\documentclass[12pt,preprint]{aastex}

\slugcomment{Submitted to ApJL: 2002--06--14; accepted: 2002--06--24}
\shorttitle{Ultraviolet $^{12}$C$^{17}$O toward X Per}
\shortauthors{Sheffer, Lambert, \& Federman}

\begin{document}
\title{Ultraviolet Detection of Interstellar $^{12}$C$^{17}$O and the CO Isotopomeric Ratios
toward X Persei}
\author{Yaron Sheffer\altaffilmark{1}, David L. Lambert\altaffilmark{2},
and S. R. Federman\altaffilmark{1}}
\altaffiltext{1}{Department of Physics and Astronomy, University of Toledo, Toledo, OH 43606;
ysheffer@physics.utoledo.edu, sfederm@uoft02.utoledo.edu.}
\altaffiltext{2}{Department of Astronomy, University of Texas, Austin, TX 78712;
dll@astro.as.utexas.edu.}

\begin{abstract}
We report the detection of fully resolved absorption lines of {\it A$-$X} bands
from interstellar $^{12}$C$^{17}$O and $^{12}$C$^{18}$O, through high-resolution spectroscopy
of \objectname{X Per} with the Space Telescope Imaging Spectrograph\footnotemark[3].
The first ultraviolet measurement of an interstellar $^{12}$C$^{17}$O column density shows that
its isotopomeric ratio is $^{12}$C$^{16}$O/$^{12}$C$^{17}$O = 8700 $\pm$ 3600. Simultaneously,
the second ultraviolet detection of interstellar $^{12}$C$^{18}$O establishes its isotopomeric ratio
at 3000 $\pm$ 600. These ratios are about five times higher than local ambient oxygen isotopic
ratios in the ISM. Such severe fractionation of rare species shows that both $^{12}$C$^{17}$O and
$^{12}$C$^{18}$O are destroyed by photodissociation, whereas $^{12}$C$^{16}$O avoids
destruction through self-shielding. This is to be contrasted with our ratio of
$^{12}$C$^{16}$O/$^{13}$C$^{16}$O = 73 $\pm$ 12 toward X Per, which is indistinguishable from
$^{12}$C/$^{13}$C, the result of a balance between photodissociation of $^{13}$C$^{16}$O and its
preferential formation via the isotope exchange reaction between CO and C$^+$.
\end{abstract}


\section{Introduction and Observations}

\footnotetext[3] {Based on observations obtained with the NASA/ESA {\it Hubble Space Telescope
(HST)} through the Space Telescope Science Institute, which is operated by the Association of
Universities for Research in Astronomy, Inc., under NASA contract NAS5-26555.}

Carbon monoxide is the second most abundant molecule in interstellar clouds after
H$_2$, with readily observable electronic transitions in the vacuum ultraviolet (VUV), vibrational
bands in the infrared, and pure rotational lines in the mm-wave regime. The isotopic varieties
of CO are used to constrain models of star formation, chemical networks, and
stellar evolution. The first VUV detection of rotationally unresolved fourth-positive
($A~^1\Pi$$-$$X~^1\Sigma^+$) absorption bands from interstellar $^{12}$C$^{16}$O and
$^{13}$C$^{16}$O was reported by \citet{ss}--see \citet{mn} for
an extensive review of the fourth-positive bands.
However, most measurements of various CO species come from radio observations
of molecular clouds with substantial total column densities ($N$) of CO.
\citet{p72} reported the first mm observations of $^{12}$C$^{16}$O and $^{13}$C$^{16}$O
in dark clouds. Rarer CO varieties were reported later: $^{12}$C$^{18}$O by \citet{mms},
$^{12}$C$^{17}$O by \citet{d77}, and $^{13}$C$^{18}$O by \citet{l80}.
The ``final'' milestone for radio CO was recently reached by \citet{b01}, who reported
the detection of the rarest stable CO isotopomer, $^{13}$C$^{17}$O, in the $\rho$ Oph
molecular cloud. In the VUV, absorption from CO is sought, but such observations sample
substantially smaller $N$(CO) than are commonly observed in the radio because ultraviolet
extinction limits the number of suitable targets behind molecular clouds. Individual {\it A$-$X}
rotational lines of interstellar absorption bands have been fully resolved
and measured only for $^{12}$C$^{16}$O and $^{13}$C$^{16}$O, using the echelle grating of
the Goddard High-Resolution Spectrograph (GHRS) on board the $HST$ \citep{s91,s92}.
A previous detection with a GHRS first-order grating of rotationally
unresolved $^{12}$C$^{18}$O was reported by \citet{l94} toward $\zeta$ Oph.

In this Letter
we report the first VUV detection of interstellar absorption from $^{12}$C$^{17}$O bands,
and present the first unblended measurements of its sibling, $^{12}$C$^{18}$O.
These observations were made toward X Per (HD 24534), using grating E140H of the Space Telescope
Imaging Spectrograph (STIS) for data sets o64812010$-$030 and o64813010$-$020.
The star was observed through the smallest aperture (0.1\arcsec~$\times$ 0.025\arcsec,
the ``Jenkins slit''), providing the highest $HST$ resolving power of
$\lambda/\Delta\lambda$ = 200,000 over $\lambda$ = 1316 to 1517 \AA. We modeled the data using
our unpublished spectrum synthesis code, ISMOD, which is based on the line transfer equations
given by \citet{bvd}. For more details of data extraction and modeling, see \citet{sfl}.
In the next section we present CO column densities along the X Per line
of sight, and in \S 3 we compare CO isotopomeric ratios with ambient carbon and oxygen
isotopic ratios in the context of theoretical models of translucent clouds.

\section{Column Densities of CO Isotopomers}

With the help of optically thin intersystem bands of CO, we derived
$N$($^{12}$C$^{16}$O) = 1.41 ($\pm$ 0.22) $\times$ 10$^{16}$ cm$^{-2}$
toward X Per from the same data sets used here, i.e.,
a column density 5.5 $\pm$ 0.9 times higher than toward $\zeta$ Oph (Sheffer et al. 2002).
The column density of the $^{13}$C$^{16}$O isotopomer was modeled in the same study to be
1.94 ($\pm$ 0.08) $\times$ 10$^{14}$ cm$^{-2}$ by analysis of its weak {\it A$-$X} (8$-$0) band.
Although the column densities of the rarer isotopomers are derived below by using the same
4-cloud model, the resulting abundances of these species are sensitive only to the total
equivalent width ($W_\lambda$) of each band, because the optical depth ($\tau$) at line center
is less than 1.

Rotational structures of four $^{12}$C$^{18}$O {\it A$-$X} bands are now resolved:
see (2$-$0), (3$-$0), (4$-$0), and (5$-$0) in Figure 1. Table 1 lists their laboratory
wavelengths computed from term values given by \citet{b97}, who measured VUV emission
line positions for $v^\prime$= 0 to 9 at $R$ = 130,000. The match of laboratory wavelengths
with observed positions of X Per absorption lines is very good, i.e., they are consistent
with the radial velocity of CO along this line of sight. Electric dipole oscillator
strengths ($f$-values) are normally assumed to be identical for all CO isotopomers.
Therefore, $N$($^{12}$C$^{18}$O) was derived by fitting the position and strength of observed
bands using published $f$-values of $^{12}$C$^{16}$O \citep{ccb}. For the
group of four {\it A$-$X} bands modeled here together, their average $f$-value taken from
Chan et al. (1993) is only 3.5\% smaller than an average based on $f$-values recommended by 
\citet{e99}. Values of band $W_\lambda$ and $\tau$[$R$(0)] are also
listed in Table 1 to show that these rare species indeed have
optically thin lines. All four bands were fitted simultaneously,
to increase the ``signal-to-noise ratio'' of the model. In a complementary manner,
individual band fits are used to derive formal error bars for modeled parameters.

The column density of $^{12}$C$^{18}$O is found to be 4.69 ($\pm$ 0.66) $\times$ 10$^{12}$ cm$^{-2}$.
Another free fitting parameter, the ground state excitation temperature
($T_{J^\prime}$$_{^\prime > 0}$,$_{J^\prime}$$_{^\prime = 0}$),
is consistent with very low rotational temperatures usually found
in the VUV for CO in diffuse clouds: $T_{1,0}$($^{12}$C$^{18}$O) = 3.4 ($\pm$ 1.0) K. For comparison,
we find $T_{1,0}$ = 3.3 K and 3.7 K from the {\it A$-$X} (8$-$0) bands of $^{12}$C$^{16}$O and
$^{13}$C$^{16}$O, respectively, toward X Per. For the $J^\prime$$^\prime$ = 2 rotational level,
we provide a 2 $\sigma$ upper limit, $T_{2,0}$ $\leq$ 6.2 K.
 
Identical steps of observational reductions and data modeling were followed with
spectral segments of $^{12}$C$^{17}$O bands, although rest wavelengths had to be computed
indirectly from $^{12}$C$^{16}$O molecular constants, see Table 1. Figure 1 presents
fits of the four {\it A$-$X} bands of $^{12}$C$^{17}$O: (2$-$0), (3$-$0), (4$-$0), and (5$-$0).
The measured column density of $^{12}$C$^{17}$O is 1.63 ($\pm$ 0.62) $\times$ 10$^{12}$ cm$^{-2}$.
A reliable determination of $T_{1,0}$ was possible only from the (3$-$0) band,
on account of the extreme weakness of the $^{12}$C$^{17}$O $J^\prime$$^\prime$ $>$ 0 lines.
Thus $T_{1,0}$($^{12}$C$^{17}$O) = 3.1 ($\pm$ 2.0) K, where twice the uncertainty from the four-band
C$^{18}$O model is adopted for this one-band result.
Note that the $X ^1\Sigma^+$ (ground state) rotational levels of $^{12}$C$^{17}$O are
intrinsically split into hyperfine components, observable as radio line separations of
$\sim$ 3 km s$^{-1}$ or less (see Figure 1 of Bensch et al. 2001). Furthermore, since
$A~^1\Pi$ hyperfine splittings should be much smaller than ground state splittings,  
hyperfine separations for VUV transitions are of the order of 10$^{-4}$ km s$^{-1}$, which we
ignore here.

\section{CO Isotopomeric Ratios toward X Per}

As described in the last section, column densities for four CO isotopomers toward X Per are now
available from STIS spectra of {\it A$-$X} bands. Note that VUV absorptions are confined
to the narrow beam subtended by the background star, whereas radiotelescopes that are used for
emission measurements have much larger beam sizes.
Of particular diagnostic value are the isotopomeric column density ratios and their departures
from ambient isotopic ratios. In Table 2 we list CO isotopomeric ratios derived from our data,
as well as carbon and oxygen isotopic ratios that were measured for local
molecular clouds and are taken from reviews by \citet{wr} and \citet{w99}.
The galactocentric distance of X Per is 9\% larger than that of the Sun; we assume that the
material along its sight line is represented by local ISM abundances.
The level of isotopic fractionation is given by $F$ $\neq$ 1, where $F$ is defined as
the isotopomeric ratio normalized by the corresponding isotopic ratio. From Table 2 we get
$F$(12,13) $\equiv$ ($^{12}$C$^{16}$O/$^{13}$C$^{16}$O)/($^{12}$C/$^{13}$C) = 1.0 $\pm$ 0.2,
$F$(16,18) = 5.4 $\pm$ 1.2, $F$(16,17) = 4.6 $\pm$ 1.9, and $F$(18,17) = 0.9 $\pm$ 0.4.
Relative to $^{12}$C$^{16}$O, the $^{13}$C$^{16}$O isotopomer is unfractionated.
Both $^{12}$C$^{18}$O and $^{12}$C$^{17}$O are quite severely fractionated
but by similar factors such that $F$(18,17) is consistent with unity within the errors of
measurement.

The line of sight through the translucent \citep[$A_V$ $\approx$ 1.44 mag]{s98} cloud
toward X Per harbors a total
CO column density of 1.43 $\times$ 10$^{16}$ cm$^{-2}$, which is one of the larger values
observed in the VUV to date, but certainly is much smaller (by 1 to 3 orders of magnitude)
than column densities commonly observed via radio emission in molecular clouds.
CO photochemistry in diffuse and translucent clouds has been studied theoretically in many papers,
e.g., \citet{vdb} and \citet{wbv}. The equilibrium CO abundance
is dependent in large part on photodissociation driven by absorption of ultraviolet photons
in bound-bound transitions to excited states from which predissociation occurs. Multiple electronic
transitions are involved in the dissociation. Calculation of the photodissociation rate depends,
of course, on the adopted interstellar radiation field, $I_{\rm UV}$, and the penetration of the
ultraviolet photons into a cloud. This latter process is affected by the ultraviolet opacity
of the embedded dust grains and by $\tau$ in the CO ``photodissociating'' lines
and in overlapping H$_2$ and H Lyman lines. When $\tau$ of such key lines is large, CO
molecules become shielded against $I_{\rm UV}$. Self-shielding may occur for transitions
not overlapped by hydrogen absorption.

All CO isotopomers are subject to photodissociation, the rate of which, $\Gamma_i$ (where $i$
stands for an isotopomeric flavor), decreases into the cloud as molecular abundances and their
opacities (and that from grains) build up. As a natural outcome of the preponderance of $^{12}$C
and $^{16}$O among all
carbon and oxygen isotopes, the primary isotopomer of CO is subject to greater self-shielding,
resulting in a steeper gradient of growing abundance into the cloud. For unshielded CO molecules,
e.g., at the edge of a cloud, the photodissociation rates of the various isotopomers are
at their maximum values and effectively identical.
Owing to wavelength shifts between the ``photodissociating'' transitions of
different isotopomers, the dominant molecule $^{12}$C$^{16}$O cannot fully self-shield the less
abundant isotopomers. Indeed, Warin et al. (1996) describe the shielding of rare isotopomers
by $^{12}$C$^{16}$O as ``not efficient''. Calculations show that $^{12}$C$^{16}$O can be shielded
from the ultraviolet over a large portion of a diffuse or translucent cloud but trace species
like $^{12}$C$^{18}$O are almost entirely unshielded. Therefore, the faster build up of
$^{12}$C$^{16}$O
with depth into the cloud is predicted to be manifested in increased fractionation because,
e.g., $F$(16,18) = $\Gamma_{1218}$/$\Gamma_{1216}$ $>$ 1.

In diffuse and translucent clouds, carbon is predominantly ionized \citep{sfm}. Consequently,
the behavior of $F$(12,13) is compounded by the isotope exchange reaction \citep{wah},
$^{13}$C$^+$ + $^{12}$C$^{16}$O $\leftrightarrow$ $^{13}$C$^{16}$O + $^{12}$C$^+$ + $\Delta$E(= 34.6 K),
which is very competitive with the photodissociation rate $\Gamma_{1316}$ because its rate
coefficient $k_{\rm f}$ is 6.8 $\times$ 10$^{-10}$ cm$^3$ s$^{-1}$
\citep[at $T$ = 80 K]{sa}, and which is able, therefore, to enhance the
abundance of $^{13}$C-containing isotopomers at the expense of $^{12}$C-containing isotopomers.
(An analogous isotope exchange reaction involving oxygen ions, working to reduce both
$F$(16,18) and $F$(16,17), is inhibited by the fact that oxygen is predominantly neutral
in these environments.) The outcome for $^{13}$C$^{16}$O is a compromise between the two isotope
selective processes, photodissociation and isotope exchange. Indeed,
published photochemical models (see more below) do show that $F$(12,13) is expected to be
$\sim$1 in the outer cloud envelope even though $\Gamma_{1316}$/$\Gamma_{1216}$ $>$ 1,
and remain $\sim$1 until the interior parts of the cloud are reached, where photodissociation
suffers great extinction and where CO becomes the most abundant form of carbon instead of C$^+$,
thus extinguishing the isotope exchange reaction.

Qualitatively, our measured isotopomeric ratios fit the above description of CO photochemistry
in translucent clouds. First, the $^{12}$C$^{16}$O/$^{13}$C$^{16}$O ratio is equal to the
ambient carbon isotopic ratio. Since $\Gamma_{1316}$ is expected to be significantly higher
than $\Gamma_{1216}$, the isotope exchange reaction must be operating in the probed regions
of the clouds to effectively reduce $F$(12,13) from $\approx$5.4, the value of $F$(16,18), to 1.
(The similarity of $\Gamma_{1316}$ and $\Gamma_{1218}$ is predicted by the similarity of the
shielding functions of the two isotopomers, as can be seen in Table 5 of van Dishoeck \& Black
1988.) A different interpretation, that of a fully efficient shielding for both
species, must imply that the isotope exchange reaction is not operating. Such a scenario is
possible only for a warm cloud with $T_{\rm kin}$ $\gtrapprox$ 160 K, a value much higher than
20 K, as was inferred from a chemical model of the X Per line of sight \citep{f94}. 
Second, both trace species $^{12}$C$^{17}$O and $^{12}$C$^{18}$O are underabundant relative to
$^{12}$C$^{16}$O by a factor of $\sim$5. This suggests that neither trace species is significantly
shielded. An inspection of the van Dishoeck \& Black (1988) translucent cloud models, especially
using their Fig. 11 of cumulative column densities ratios from model T6, shows that $F$(16,18)
does, indeed, climb up to $\sim$5 some 1 pc into the cloud, where $A_V$ $\sim$ 1.2 mag from
cloud edge. Simultaneously, the value of $F$(12,13) remains close to 1 throughout the cloud,
because both isotope selective processes very nearly cancel each other.

Unfortunately, CO column densities obtained from Table 5 of van Dishoek \& Black (1988)
do not agree with our observed values, in the sense that in order to reproduce
observed isotopomeric ratios, one has to pick those models with $N$(CO) much higher than observed.
Specifically, models T2 and H3 have values of $A_V$, $N$(H$_2$), $N$(C$^+$), and
$N$($^{12}$C$^{16}$O) that match the X Per line of sight. However, the same models display values
for $F$(12,13) and $F$(16,18) that are some 50\% smaller than our observed values.
On the other hand, whereas our isotopomeric ratios are matched by
CO ratios from models T4, H4, and H5, the corresponding modeled CO column densities for all
isotopomers are 10 to 40 times higher than observed values. In fact, the high-$N$(CO) discrepancy
is highlighted by the persistent agreement between modeled and observed $N$(C$^+$) values.
Part of the discrepancy may be due
to the X Per line of sight passing through the cloud envelope, but not through the cloud center.
Such a scenario has been explored by \citet{krp}, who showed that
CO (but not C$^+$) is especially sensitive to model geometry. In particular, for the same total
extinction, off-center optical paths through a spherical cloud can generate $N$(CO) values lower
by 1 to 2 orders of magnitude than values from a path that is perpendicular to a plane-parallel
slab, the likes of which were employed by van Dishoek \& Black (1988) and by Warin et al. (1996).
Interestingly, Snow et al. (1998) suggested that the cloud toward X Per might be a dispersing
remnant of a denser molecular cloud. Besides not resembling a plane-parallel slab, this
cloud may also pose modeling difficulties in terms of unsatisfied steady-state assumptions.
 
Likewise, it is possible to find partial agreement between the single, $A_V$ = 4.34 mag model of
a translucent cloud from Warin et al. (1996) and our results. As abundances of all CO species
increase into the cloud owing to enhanced UV shielding (see their Figure 7), both
$^{12}$C$^{16}$O and $^{13}$C$^{16}$O first reach a plateau at $A_V$ $\sim$1.0, where their
mutual ratio is $\sim$70, or $F$(12,13) $\sim$0.8, since their input carbon isotopic ratio is 90.
Furthermore, at that very point into the cloud, the
ratio of $^{12}$C$^{16}$O to $^{12}$C$^{18}$O stands at $\sim$2000, i.e., $F$(16,18) $\sim$4.
But in a similar fashion to the models of van Dishoeck \& Black (1988), predicted $N$(CO)
values are more than 10 times higher than those observed toward X Per. Warin et al. (1996) also
modeled individual rotational level populations for the CO isotopomers,
since level photodissociation rates are coupled to ground state excitation temperatures. Modeled
$T_{J^\prime}$$_{^\prime > 0}$,$_{J^\prime}$$_{^\prime = 0}$ values
are higher than our observed values, with $T_{1,0}$ $\sim$ 0.1 $\times$ $T_{\rm kin}$ $\sim$ 5 K.
One possible remedy is to assume that the same ratio of $T_{1,0}$ over $T_{\rm kin}$ applies
to the X Per clouds. In that case, the observed $T_{1,0}$ value of $\approx$ 3.5 K may indicate
that $T_{\rm kin}$ is $\approx$ 35 K toward X Per, rather than 20 K. Indeed, \citet{k00} derived
$T_{\rm kin}$ = 45 ($\pm$ 15) K from observations of C$_2$ along this line of sight. 

\section{Concluding Remarks}

The line of sight toward X Per has provided us with a rich spectrum of {\it A$-$X} bands of CO,
allowing the first VUV detection of the rare isotopomer $^{12}$C$^{17}$O and the first
rotationally-resolved views of both $^{12}$C$^{18}$O and $^{12}$C$^{17}$O. These detections
were made possible by the superb qualities of STIS as a VUV spectrometer. Toward X Per we find that
$^{13}$C$^{16}$O is unfractionated with respect to $^{12}$C$^{16}$O owing to a balance between
the rates of photodissociation and of the isotope exchange reaction. On the other hand, the lack
of an isotope exchange reaction in the case of oxygen isotopes renders both $^{12}$C$^{18}$O
and $^{12}$C$^{17}$O strongly fractionated and destroyed at the 80\% level with respect
to the strongly shielded $^{12}$C$^{16}$O. As was described above, similar isotopomeric ratios
are to be found within published results from theoretical models. However, a large gap remains
between observed and modeled column densities for CO. 

With the detection of $^{12}$C$^{17}$O toward X Per, observers of interstellar absorption lines are
left with the last two stable CO isotopomers that have yet to be detected in the ultraviolet,
namely, $^{13}$C$^{18}$O and $^{13}$C$^{17}$O. These are very challenging tasks at best, as the
respective intrinsic abundances are expected to be \case{1}{24} and \case{1}{70} the abundance of
$^{12}$C$^{17}$O. Whereas $^{13}$C$^{18}$O was included in models of translucent clouds,
not even a coarse grid of theoretical models exists for $^{12}$C$^{17}$O in translucent clouds.
Hopefully, the new observations of this species as reported here and of $^{13}$C$^{17}$O
by Bensch et al. (2001) will provide an incentive for
the inclusion of $^{17}$O-bearing molecules in computer simulations. There is also
a need for laboratory measurements, since the {\it A$-$X} bands of $^{12}$C$^{17}$O currently
lack rigorous wavelength and perturbation analyses.

\acknowledgments

We thank an anonymous referee for very rapid handling of this manuscript.
The research presented here was supported in part by a NASA grant for $HST$ program
GO-08622 and NASA grant NAG5-4957 to the University of Toledo.

\begin{deluxetable}{llllllllll}
\tablecolumns{10}
\tablewidth{0pt}
\tabletypesize{\scriptsize}
\tablecaption{Rare Species of CO toward X Per}
\startdata
\hline \hline\\
 & \multicolumn{4}{c}{$^{12}$C$^{18}$O} & & \multicolumn{4}{c}{$^{12}$C$^{17}$O}\\
\cline{2-5} \cline{7-10}\\
{\it A$-$X}  &(2$-$0)  &(3$-$0)  &(4$-$0)  &(5$-$0)&  &(2$-$0)  &(3$-$0)  &(4$-$0)  &(5$-$0)\\
$\lambda_0$[$R$(1)] (\AA)	&1478.845  &1449.248  &1421.487 &1395.416 & &1478.219  &1448.332  &1420.320  &1394.029\\
$\lambda_0$[$R$(0)] (\AA)	&1478.895  &1449.294  &1421.530 &1395.455 & &1478.270  &1448.379  &1420.363  &1394.069\\
$\lambda_0$[$Q$(1)] (\AA)	&1478.975  &1449.371  &1421.604 &1395.526 & &1478.352  &1448.458  &1420.439  &1394.142\\
$v_{\rm helio}$ (km s$^{-1}$)  &14.8  &15.0  &14.4  &15.9 & &13.8  &13.5  &13.4  &13.5\\
$W_\lambda$ (m\AA)  &3.14 $\pm$ 0.26  &2.65 $\pm$ 0.17  &1.84 $\pm$ 0.23  &1.10 $\pm$ 0.14 & &1.22 $\pm$ 0.21  &1.02 $\pm$ 0.14  &0.69 $\pm$ 0.17  &0.40 $\pm$ 0.17\\
$\tau$[$R$(0)]  &0.6  &0.5  &0.4  &0.2 & &0.2  &0.2  &0.1   &0.1\\
\enddata
\tablecomments{Laboratory wavenumbers for $^{12}$C$^{18}$O were calculated from the term values
of Beaty et al. (1997). Wavenumbers for $^{12}$C$^{17}$O were computed from Dunham coefficients for
$^{12}$C$^{16}$O (corrected by the proper reduced-mass factors) and then shifted to correct for
$^{12}$C$^{18}$O wavenumber differences between our inferred values and those of Beaty et al. (1997).
Although such shifts were not larger than 0.16 cm$^{-1}$ (3 m\AA), the apparent velocity difference
of 1.4 km s$^{-1}$ (7 m\AA) between the two species indicates that a possible systematic shift may
be affecting our computed $^{12}$C$^{17}$O wavenumbers.}
\end{deluxetable}

\begin{deluxetable}{lll}
\tablecolumns{10}
\tablewidth{0pt}
\tablecaption{Isotopic Comparisons toward X Per}
\startdata
\hline \hline\\
\colhead{Isotopomeric Ratios} & & \colhead{Local Isotopic Ratios}\\
\cline{1-1} \cline{3-3}\\
$\frac{N({\rm ^{12}CO})}{N({\rm ^{13}CO})} = 73 \pm 12$    & & $\frac{{\rm ^{12}C}}{{\rm ^{13}C}} = 70 \pm 7$ \\
$\frac{N({\rm C^{16}O})}{N({\rm C^{18}O})} = 3000 \pm 600$ & & $\frac{{\rm ^{16}O}}{{\rm ^{18}O}} = 560 \pm 25$ \\
$\frac{N({\rm C^{16}O})}{N({\rm C^{17}O})} = 8700 \pm 3600$& & $\frac{{\rm ^{16}O}}{{\rm ^{17}O}} = 1900 \pm 200$ \\
$\frac{N({\rm C^{18}O})}{N({\rm C^{17}O})} = 2.9 \pm 1.2$  & & $\frac{{\rm ^{18}O}}{{\rm ^{17}O}} = 3.4 \pm 0.2$ \\
\enddata
\end{deluxetable}

\begin{figure}
\epsscale{0.88}
\plotone{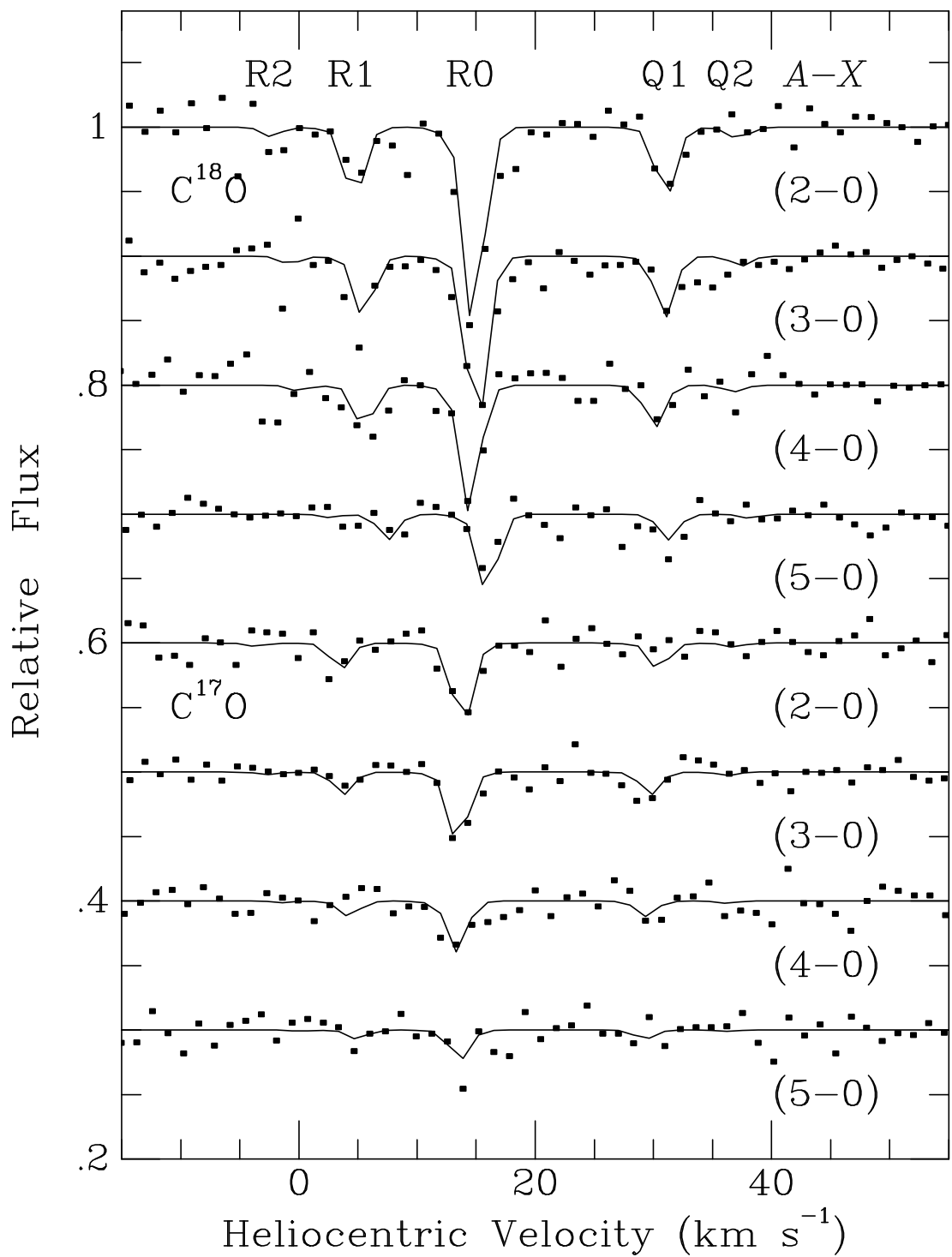}
\caption{Montage of four {\it A$-$X} bands of $^{12}$C$^{18}$O and four {\it A$-$X} bands of
$^{12}$C$^{17}$O. STIS data toward X Per are shown as filled squares, while global models are
shown by solid lines. Both data and models have been rebinned by two for clarity. The heliocentric
radial velocity is shown for the $R$(0) line of all bands, which are shifted by $-$0.1 in
continuum units.}
\end{figure}


\begin{thebibliography}{}
\bibitem[Beaty et al.(1997)]{b97} Beaty, L. M., Braun, V. D., Huber, K. P., \& Le Floch, A. C. 1997, \apjs, 109, 269
\bibitem[Bensch et al.(2001)]{b01} Bensch, F., Pak, I., Wouterloot, J. G. A., Klapper, G., \& Winnewisser, G. 2001, \apj, 562, L185
\bibitem[Black \& van Dishoeck(1988)]{bvd} Black, J. H., \& van Dishoeck, E. F. 1988, \apj, 331, 986
\bibitem[Chan, Cooper, \& Brion(1993)]{ccb} Chan, W. F., Cooper, G., \& Brion, C. E. 1993, Chem. Phys., 170, 123
\bibitem[Dickman et al.(1977)]{d77} Dickman, R. L., Langer, W. D., McCutcheon, W. H., \& Shuter, W. L. H. 1977, in CNO Isotopes in Astrophysics, ed. J. Adouze (Dordrecht: Reidel), 95
\bibitem[Eidelsberg et al.(1999)]{e99} Eidelsberg, M., Jolly, A., Lemaire, J. L., Tchang-Brillet, W.-\"{U}. L., Breton, J., \& Rostas, F. 1999, \aap, 346, 705
\bibitem[Federman et al.(1994)]{f94} Federman, S. R., Strom, C. J., Lambert, D. L., Cardelli, J. A., Smith, V. V., \& Joseph, C. L. 1994, \apj, 424, 772
\bibitem[Kaczmarczyk(2000)]{k00} Kaczmarczyk, G. 2000, Acta Astron., 50, 151
\bibitem[Kopp, Roueff, \& Pineau des For\^{e}ts(2000)]{krp} Kopp, M., Roueff, E., \& Pineau des For\^{e}ts, G. 2000, \mnras, 315, 37
\bibitem[Lambert et al.(1994)]{l94} Lambert, D. L., Sheffer, Y., Gilliland, R. L., \& Federman, S. R. 1994, \apj, 420, 756
\bibitem[Langer et al.(1980)]{l80} Langer, W. D., Goldsmith, P. F., Carlson, E. R., \& Wilson, R. W. 1980, \apj, 235, L39
\bibitem[Mahoney, McCutcheon, \& Shuter(1976)]{mms} Mahoney, M. J., McCutcheon, W. H., \& Shuter, W. L. H. 1976, \aj, 81, 508
\bibitem[Morton \& Noreau(1994)]{mn} Morton, D. C., \& Noreau, L. 1994, \apjs, 95, 301
\bibitem[Penzias et al.(1972)]{p72} Penzias, A. A., Solomon, P. M., Jefferts, K. B., \& Wilson, R. W. 1972, \apj, 174, L43
\bibitem[Sheffer, Federman, \& Lambert(2002)]{sfl} Sheffer, Y., Federman, S. R., \& Lambert, D. L. 2002, \apj, 572, L95
\bibitem[Sheffer et al.(1992)]{s92} Sheffer, Y., Federman, S. R., Lambert, D. L., \& Cardelli, J. A. 1992, \apj, 397, 482
\bibitem[Smith \& Stecher(1971)]{ss} Smith, A. M., \& Stecher, T. P. 1971, \apj, 164, L43
\bibitem[Smith et al.(1991)]{s91} Smith, A. M., et al. 1991, \apj, 377, L61
\bibitem[Smith \& Adams(1980)]{sa} Smith, D., \& Adams, N. G. 1980, \apj, 242, 424
\bibitem[Snow et al.(1998)]{s98} Snow, T. P., Hanson, M. M., Black, J. H., van Dishoeck, E. F., Crutcher, R. M., \& Lutz, B. L. 1998, \apj, 496, L113
\bibitem[Sofia, Fitzpatrick, \& Meyer(1998)]{sfm} Sofia, U. J., Fitzpatrick, E. L., \& Meyer, D.M. 1998, \apj, 504, L47
\bibitem[van Dishoeck \& Black(1988)]{vdb} van Dishoeck, E. F., \& Black, J. H., 1988, \apj, 334, 771
\bibitem[Warin, Benayoun, \& Viala(1996)]{wbv} Warin, S., Benayoun, J. J., \& Viala, Y. P.  1996, \aap, 308, 535
\bibitem[Watson, Anicich, \& Huntress(1976)]{wah} Watson, W. D., Anicich, V. G., \& Huntress, W. T. Jr. 1976, \apj, 205, L165
\bibitem[Wilson(1999)]{w99} Wilson, T. L. 1999, Rep. Prog. Phys. 62, 143
\bibitem[Wilson \& Rood(1994)]{wr} Wilson, T. L., \& Rood, R. T. 1994, ARA\&A, 32, 191
\end{thebibliography}
\end{document}